\documentclass[twocolumn,showpacs,prc,amsmath,amssymb,superscriptaddress]{revtex4}

\usepackage{graphicx}
\usepackage{dcolumn}
\usepackage{bm}
\usepackage{epsfig}

\begin{document}

\preprint{APS/123-QED}

\title{Production of medium-mass neutron-rich nuclei in reactions induced by $^{136}$Xe projectiles at 1 A GeV on a beryllium target}

\author{J.~Benlliure}
\email{j.benlliure@usc.es}
\affiliation{Universidade de Santiago de Compostela, E-15782 Santiago de 
Compostela, Spain}
\author{M.Fern\'andez-Ord\'o\~nez}
\altaffiliation[Present address: ]{CIEMAT, Avd. Complutense, 28000 Madrid, Spain}
\affiliation{Universidade de Santiago de Compostela, E-15782 Santiago de 
Compostela, Spain} 
\author{L.~Audouin}
\affiliation{IPN Orsay, Universit\'e Paris-Sud 11, CNRS/IN2P3, F-91406 Orsay cedex, France}
\author{A.~Boudard}
\affiliation{DAPNIA/SPhN, DSM-CEA, F-91191 Gif-sur-Yvette cedex, France}
\author{E.~Casarejos}
\affiliation{Universidade de Santiago de Compostela, E-15782 Santiago de 
Compostela, Spain} 
\author{J.E.~Ducret}
\affiliation{DAPNIA/SPhN, DSM-CEA, F-91191 Gif-sur-Yvette cedex, France}
\author{T.~Enqvist}
\altaffiliation[Present address: ]{CUPP project, P.O. Box 22, 86801 Pyh\"asalmi, Finland}
\affiliation{GSI, Planckstr.~1, D-64291~Darmstadt, Germany}
\author{A.~Heinz}
\altaffiliation[Present address: ]{WNSL, Yale University, New Haven, Connecticut 06511, USA}
\affiliation{GSI, Planckstr.~1, D-64291~Darmstadt, Germany}
\author{D.~Henzlova}
\altaffiliation[Present address: ]{NSCL, Michigan State University, East Lansing, Michigan 48824, USA}
\affiliation{GSI, Planckstr.~1, D-64291~Darmstadt, Germany}
\author{V.~Henzl}
\altaffiliation[Present address: ]{NSCL, Michigan State University, East Lansing, Michigan 48824, USA}
\affiliation{GSI, Planckstr.~1, D-64291~Darmstadt, Germany}
\author{A.~Kelic}
\affiliation{GSI, Planckstr.~1, D-64291~Darmstadt, Germany}
\author{S.~Leray}
\affiliation{DAPNIA/SPhN, DSM-CEA, F-91191 Gif-sur-Yvette cedex, France}
\author{P.~Napolitani}
\altaffiliation[Present address: ]{LPC Caen, ENSICAEN, Universit\'e de Caen, CNRS/IN2P3, F-14050 Caen cedex 4, France}
\affiliation{IPN Orsay, Universit\'e Paris-Sud 11, CNRS/IN2P3, F-91406 Orsay cedex, France}
\affiliation{GSI, Planckstr.~1, D-64291~Darmstadt, Germany}
\author{J.~Pereira}
\altaffiliation[Present address: ]{NSCL, Michigan State University, East Lansing, Michigan 48824, USA}
\affiliation{Universidade de Santiago de Compostela, E-15782 Santiago de 
Compostela, Spain} 
\author{F.~Rejmund}
\altaffiliation[Present address: ]{GANIL, CEA/DSM-CNRS/IN2P3, BP 55027, F-14076 Caen cedex, France}
\affiliation{IPN Orsay, Universit\'e Paris-Sud 11, CNRS/IN2P3, F-91406 Orsay cedex, France}
\author{M.V.~Ricciardi}
\affiliation{GSI, Planckstr.~1, D-64291~Darmstadt, Germany}
\author{K.-H.~Schmidt}
\affiliation{GSI, Planckstr.~1, D-64291~Darmstadt, Germany}
\author{C.~Schmitt}
\altaffiliation[Present address: ]{Universit\'e Lyon I, CNRS/IN2P3, IPNL, F-69622 Villeurbanne cedex, France}
\affiliation{GSI, Planckstr.~1, D-64291~Darmstadt, Germany}
\author{C.~St\'ephan}
\affiliation{IPN Orsay, Universit\'e Paris-Sud 11, CNRS/IN2P3, F-91406 Orsay cedex, France}
\author{L.~Tassan-Got}
\affiliation{IPN Orsay, Universit\'e Paris-Sud 11, CNRS/IN2P3, F-91406 Orsay cedex, France}
\author{C. Volant}
\affiliation{DAPNIA/SPhN, DSM-CEA, F-91191 Gif-sur-Yvette cedex, France}
\author{C. Villagrasa}
\altaffiliation[Present address: ]{IRSN, BP17, F-92262 Fontenay-aux-Roses cedex, France}
\affiliation{DAPNIA/SPhN, DSM-CEA, F-91191 Gif-sur-Yvette cedex, France}
\author{O.~Yordanov}
\altaffiliation[Present address: ]{INRNE, 72 Tzarigradsko chausee, BG-1784 Sofia, Bulgaria}
\affiliation{GSI, Planckstr.~1, D-64291~Darmstadt, Germany}

\date{\today}

\begin{abstract}
Production cross sections of medium-mass neutron-rich nuclei obtained in the fragmentation of $^{136}$Xe projectiles at 1 A GeV have been measured with the FRagment Separator (FRS) at GSI. $^{125}$Pd was identified for the first time. The measured cross sections are compared to $^{238}$U fission yields and model calculations in order to determine the optimum reaction mechanism to extend the limits of the chart of the nuclides around the r-process waiting point at N=82.
\end{abstract}


\maketitle

\section{Introduction}

The access to medium-mass neutron-rich nuclei around N = 82 is a pre-requisite for investigating the evolution of nuclear shell structure with neutron excess and its implications on the stellar nucleosynthesis in the r-process. The quenching of the neutron shell gaps in neutron-rich nuclei is being extensively addressed in both, experimental \cite{Dil03,Wal04,Mon06,Jun07,Kay08} and theoretical works \cite{Dob96,Pea96,Ots05}. Recent contradictory results require more experimental investigations. However, the refractory nature of the nuclei of interest and the limited primary-beam intensities from existing fragmentation facilities prevents us from any sizeable extension of the present limits of known nuclei on the chart of the nuclides at N = 82. 

In the near future, new radioactive-beam facilities will offer improved possibilities for extensive experimental investigations. However, the choice of the appropriate reaction mechanism for the production of neutron-rich nuclei will be decisive for the magnitude of the attainable yields. Fission has been used successfully for producing a large variety of neutron-rich nuclei close to N = 82 both in in-flight \cite{Ber97} and in ISOL \cite{She02} facilities. However, the fission yields were found to drastically decrease for light N = 82 isotones, since the fluctuations in the charge-polarization degree of freedom are rather small \cite{Arm70}. An alternative option to fission is the fragmentation of stable $^{136}$Xe projectiles \cite{Mon06}. It is very interesting to explore how the cross sections of light neutron-rich nuclei close to N = 82 develop when using this fundamentally different mechanism in its extreme tail of cold fragmentation \cite{Ben99}, where fluctuations in the N/Z degree of freedom in the formation of projectile spectators with low excitation energies are exploited. More recently, it has been proposed to profit from the high secondary-beam intensities of some neutron-rich fission fragments available in ISOL facilities to use a two-stage reaction scheme for producing medium-mass neutron-rich nuclei \cite{Hel03}. According to this scheme, fission products from an ISOL facility could be re-accelerated and fragmented in order to produce nuclei with larger neutron excess. With this approach it is also possible to overcome the difficulties of the ISOL method to produce isotopes of refractory elements. 

In this paper we present the results of an experiment recently conducted at GSI where medium-mass neutron-rich nuclei produced in the reaction $^{136}$Xe+Be at 1 A GeV were separated and identified in-flight using the FRagment Separator (FRS). Moreover, the production cross sections of the projectile residues could be determined with high accuracy. In the first sections of the paper we present the experimental technique and the main results obtained in this experiment. Then, we validate different model calculations describing the production of residual nuclei in fragmentation reactions at relativistic energies. We conclude the paper with a comparison of the expected production yields of neutron-rich nuclei close to N=82 in fragmentation and fission reactions.

\section{Experiment}

\begin{figure}
  \begin{center}
    \leavevmode
    \epsfig{file=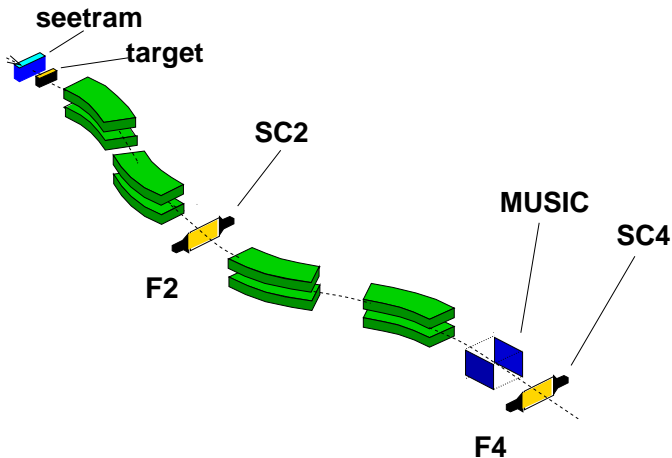,height=6cm}
\caption{Schematic layout of the FRS showing only the dipole magnetic elements and its detection system. Multiplets of quadrupoles and sextupoles are not shown in the figure.}
\label{fig_1}
  \end{center}
\end{figure}

The experiment was performed at the GSI facilities where the SIS18 synchrotron accelerated a $^{136}$Xe beam to 1 A GeV. The average intensity of the beam pulses was 10$^8$ ions, their duration 3 seconds and the total cycle 6 seconds. A secondary-electron monitor \cite{Jun96} registered continuously the beam intensity for normalisation purposes. The beam impinged onto a beryllium target with a thickness of 2.5 g/cm$^2$, located at the entrance of the FRagment Separator (FRS) \cite{Gei92}. The FRS is a zero-degree magnetic spectrometer with a resolving power of $\Delta B\rho/B\rho\approx$ 1500, a momentum acceptance of $\Delta p/p \approx$ 3\% and an angular acceptance around the central trajectory of 15 mrad \cite{Ben02}. The two symmetric stages composed of two dipoles each and several multiplets of quadrupoles and sextupoles (see Fig. \ref{fig_1}) were used in an achromatic mode with a dispersive intermediate image plane.

Reaction residues with trajectories inside the FRS acceptance for a given magnetic tuning were identified from their magnetic rigidity and time of flight. The magnetic rigidity was obtained from the measured positions of the trajectories at the intermediate and final image planes using two position-sensitive plastic scintillators. These detectors also provided the time of flight of the transmitted nuclei between these two image planes. Moreover, the atomic number of each residual nucleus was determined from its energy loss in an ionisation chamber placed at the end of the FRS and from its velocity. A detailed description of the experimental technique and data-sorting procedure can be found in Refs \cite{Tai03,Cas06,Nap07}.

\section{Results}

\begin{figure}
  \begin{center}
    \leavevmode
    \epsfig{file=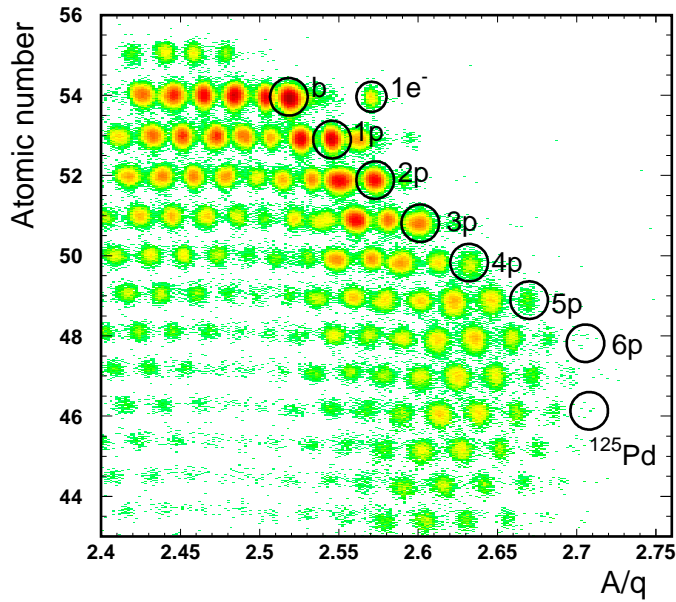,height=8cm}
\caption{Two-dimensional scatter plot of the energy loss of the nuclei transmitted through the FRS versus their A/q value obtained from the magnetic rigidity and time-of-flight measurements. This identification matrix was obtained by adding up several magnetic tunings of the FRS. The primary beam, fully stripped (b) and carrying one electron (1e$^-$), as well as the proton-loss channels (1p, 2p, ...) and $^{125}$Pd are indicated.}
\label{fig_2}
  \end{center}
\end{figure}

In figure \ref{fig_2} we present the identification matrix obtained by adding up several magnetic tunings of the FRS in the form of a two-dimensional scatter plot of the energy loss of the transmitted nuclei in the ionisation chamber versus their A/q value, obtained from the magnetic-rigidity and velocity measurements. Fig. \ref{fig_2} also illustrates the typical resolution obtained in this experiment A/$\Delta$A$\approx$ 360 and Z/$\Delta$Z$\approx$ 190.

At this projectile energy, nuclei with atomic numbers smaller than 55 are fully ionised at $\leq$ 98\%. Residual nuclei changing their atomic charge state at the intermediate image plane of the spectrometer were identified from the combined measurement of the their magnetic rigidities in both sections of the FRS and their energy loss in the ionisation chamber \cite{Ben99}. Only isotopes having three neutrons less and keeping one electron all along the FRS could contaminate the production of a given isotope at a level of 10$^{-4}$. Considering that cross sections of neutron-rich nuclei differing in three neutrons can vary at most by two orders of magnitude, the expected contamination by charge states will be of the order of few percent.

The atomic number was calibrated using the primary beam (referred as ``b'' in the figure) as reference. From this procedure, and assuming fully-stripped residues one can obtain the mass number of the nuclei using the identification matrix shown in Fig. \ref{fig_2}. In particular, in this figure we can easily identify the six first proton-removal channels (1p, 2p,...) from $^{136}$Xe ($^{135}$I, $^{134}$Te, $^{133}$Sb, $^{132}$Sn, $^{131}$In and $^{130}$Cd). The production of $^{125}$Pd is also clearly shown in this figure: this was the first time this nucleus was produced using fragmentation reactions. In a later experiment $^{125}$Pd was also identified in the fission of $^{238}$U \cite{Mot07}. 

Fig. \ref{fig_2} also shows the production of caesium isotopes as well as N=83 nuclei which have one neutron more than the projectile. All these nuclei are produced in charge-exchange reactions, as shown by A. Kelic et al. \cite{Kel04}, and will be the subject of discussion in a forthcoming publication.

The measured production yields were transformed into production cross sections by normalising to the number of incident projectiles and target thickness. Before that, the measured yields were corrected by the FRS acceptance and losses due to the dead time of the data acquisition, reactions in all layers of matter along the FRS, multiple reactions in the target and charge states of the residual nuclei. For most of the measured nuclei, no correction for the limited momentum acceptance had to be applied. The full momentum distribution was obtained by adding up the momentum-distribution segments measured in different magnetic tunings of the FRS with a difference in the magnetic values smaller than the 3\% of the momentum acceptance of the spectrometer. Using this method, the final production cross sections of the different projectile residues identified in this experiment were determined with an accuracy between 10\% and 20\%. A detailed description of the procedure can be found in Refs. \cite{Ben99,Cas06,Nap07}.

\begin{figure}
  \begin{center}
    \leavevmode
    \epsfig{file=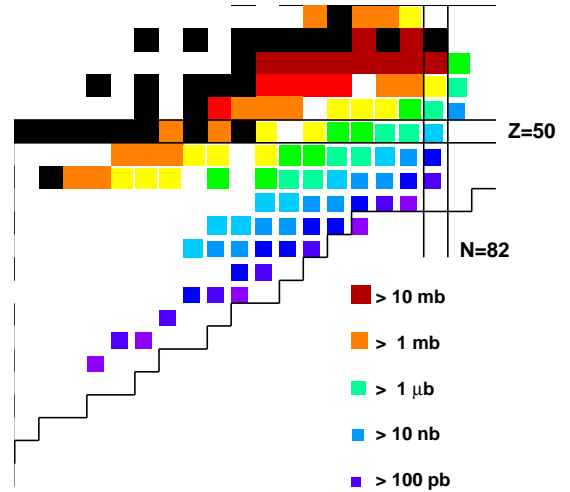,height=8cm}
\caption{Residual nuclei produced in the fragmentation of $^{136}$Xe projectiles impinging on a beryllium target at 1000 A MeV identified in this work represented on top of a chart of the nuclides. The cluster size and colour indicate the production cross section according to the legend.}
\label{fig_3}
  \end{center}
\end{figure}

Figure \ref{fig_3} summarises all projectile residues measured in this work on top of the chart of the nuclides. As it can be seen, in this measurement we covered the most neutron-rich residues produced in the fragmentation of $^{136}$Xe projectiles with production cross sections down to 100 pb.

\section{Discussion}

\subsection{Isotopic distributions of residual nuclei}

\begin{figure*}
  \begin{center}
    \leavevmode
    \epsfig{file=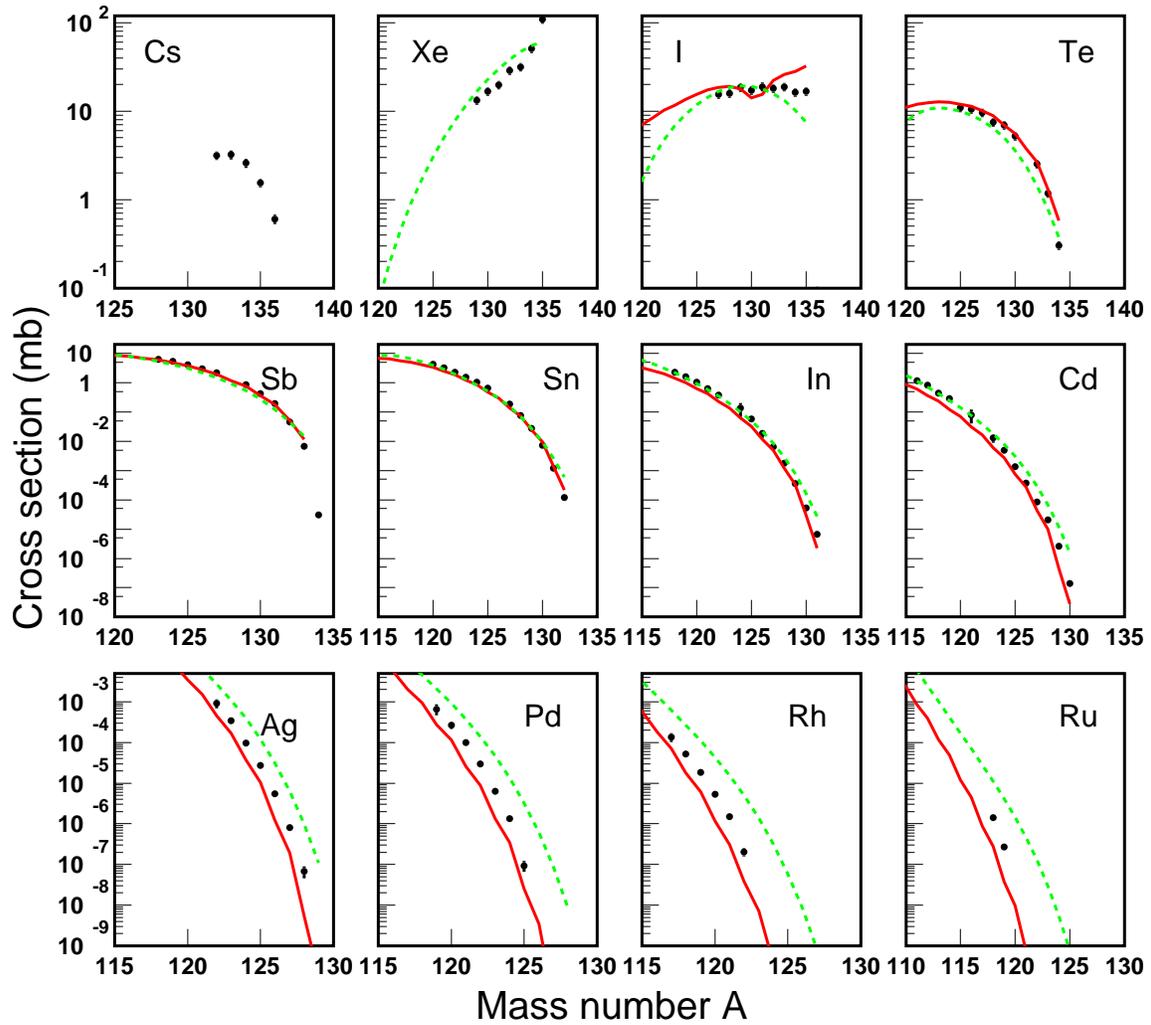,height=15cm}
\caption{Isotopic distributions of the production cross sections of projectile residues measured in this work. The measured data (solid points) are compared to the predictions obtained with two model calculations, COFRA (solid line) and EPAX (dashed line).}
\label{fig_4}
  \end{center}
\end{figure*}

The production cross sections of projectile residual nuclei measured in this work are shown in figure \ref{fig_4} in form of isotopic distributions. The error bars are visible when larger than the size of the data points. The smooth evolution of the measured cross sections, both in mass and atomic number, is considered as an indication of the high accuracy of the measurements. 

 As shown in this figure, for most of the elements produced as projectile residues, a large fraction of their isotopic distributions was measured in the present experiment, covering in particular the most neutron-rich isotopes produced with cross sections larger than 100 pb. Therefore, these data are relevant to investigate the possibilities offered by projectile-fragmentation reactions to produce medium-mass neutron-rich nuclei. 

Residual nuclei close to the projectile, isotopes of xenon and iodine, present isotopic distributions with small variations in cross section, decreasing with the neutron number. This behaviour indicates that those nuclei are produced in extremely peripheral reactions where the projectile loses at most one proton and the excitation energy gained in the reaction leads to the evaporation of few neutrons. The residual nuclei of elements lighter than iodine show a different shape in their isotopic distribution. The maximum of the distribution is located at a neutron number smaller than the one of the projectile, while the cross sections decrease with increasing neutron excess for the most neutron-rich residues. Moreover, the decrease of the production cross section for decreasing the proton number in an isotonic sequence is steeper than the decrease of the production cross section with increasing neutron number in an isotopic sequence. Indeed, for the most neutron-rich residues of silver or palladium a difference of one neutron corresponds to a variation of the cross section of one order of magnitude, while a difference of one proton corresponds to a variation of two orders of magnitude.

The production of the most neutron-rich residues is expected in reactions dominated by the cold-fragmentation process \cite{Ben99}. These are reaction channels where the projectile nucleus mostly loses protons in the interaction with the target and, at the same time, deposits only little excitation energy, allowing for the evaporation of only a few neutrons. This reaction channel leads then to the production of the most neutron-rich nuclei that can be produced in the fragmentation of a given projectile nucleus. Obviously, the production of the most neutron-rich final residues relies on large fluctuations in the proton-to-neutron ratio of the abraded nucleons and in the excitation energy gained by the projectile pre-fragment in the abrasion process. In this scenario, the most extreme case corresponds to the proton-removal channels where only protons are abraded and the excitation energy gained remains below the neutron-evaporation threshold. As already mentioned in the previous section, in this experiment we were able to identify and measure the production cross sections up to the six proton-removal channel, as depicted in Fig.\ref{fig_5}. Again this figure clearly shows how the loss of an additional proton corresponds to a reduction of almost two orders of magnitude in the final production cross section.

\subsection{Benchmarking of model calculations}

\begin{figure}
  \begin{center}
    \leavevmode
    \epsfig{file=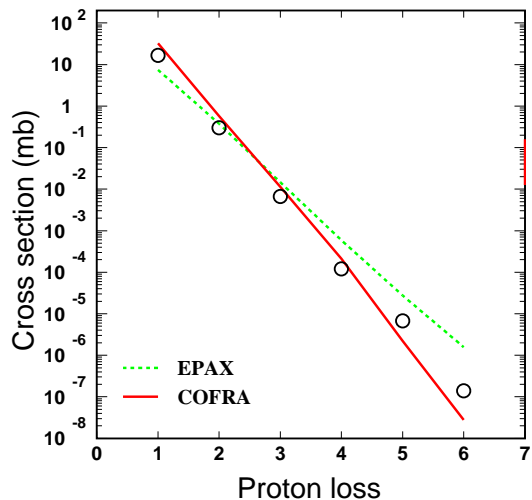,height=7cm}
\caption{Production cross sections of the proton-removal channels from $^{136}$Xe measured in this work. The data points are compared to the predictions obtained with two approaches, the COFRA model (solid line) and EPAX formula (dashed line).}
\label{fig_5}
  \end{center}
\end{figure}

The future perspectives for the production of medium-mass neutron-rich nuclei using fragmentation reactions can be investigated using model calculations which the present data validate. For this purpose we chose two different approaches. The EPAX formula \cite{Sum00}, a semi-empirical parametrisation of previously measured data and the code COFRA \cite{Ben99} a simplified version of the abrasion-ablation model ABRABLA \cite{Gai91}. COFRA includes a complete description of the abrasion process following the ideas introduced by Gaimard and Schmidt \cite{Gai91} with a reduction of the excitation energy by 10\% with respect to the description proposed in Ref. \cite{Sch93}. Then the evaporation stage is based on the statistical model using the Weisskopf prescription but considering only neutron emission as open de-excitation channel. This simplification allows for an analytical description of the evaporation process as described in \cite{Ben99} but restricts the validity of the results to those cases where the de-excitation of an excited nucleus can be described mostly by the evaporation of neutrons, as it is the case of nuclei with a large neutron excess, and in particular, those produced in cold-fragmentation reactions.  

In figures \ref{fig_4} and \ref{fig_5} we compare the results obtained with both calculations with the data measured in the present work. The EPAX formula in general provides a good description of the production cross sections of neutron-deficient fragmentation residues \cite{Caa04}. In the present case, EPAX describes rather well the production cross sections of residual nuclei not too different in mass number from the projectile. However, for residual nuclei with a large neutron excess EPAX clearly overestimates the production cross sections. This effect increases with the difference in mass number between the residual nucleus and the projectile, being close to one order of magnitude for the six proton-removal channel shown in Fig. \ref{fig_5}, and even larger for the most neutron-rich isotopes of Pd and Rh shown in Fig. \ref{fig_4}. This overestimation of the production cross sections of residual nuclei with large neutron excess and far from the projectile obtained with EPAX was already observed in previous works \cite{Ben99,Hel03,Ben07}.

The code COFRA provides a better overall description of the present data. Nevertheless, we can also identify a clear tendency to slightly under-predict the production cross sections of neutron-rich residual nuclei with a large difference in mass number with respect to the projectile, as can be seen in Figs. \ref{fig_4} and \ref{fig_5}. It should be stressed that the predictions of the COFRA code are extremely sensitive to the precise values of the neutron separation energies of the nuclei of interest. Indeed, an overestimation of few hundred keV in the neutron binding energies could explain the observed deviations from the measured cross section. A similar approach has been recently used in Ref. \cite{Tsa07} to determine the binding energies of several Cu isotopes.

\subsection{Fragmentation and fission competition for the production of medium-mass neutron-rich nuclei.}

\begin{figure*}
  \begin{center}
    \leavevmode
    \epsfig{file=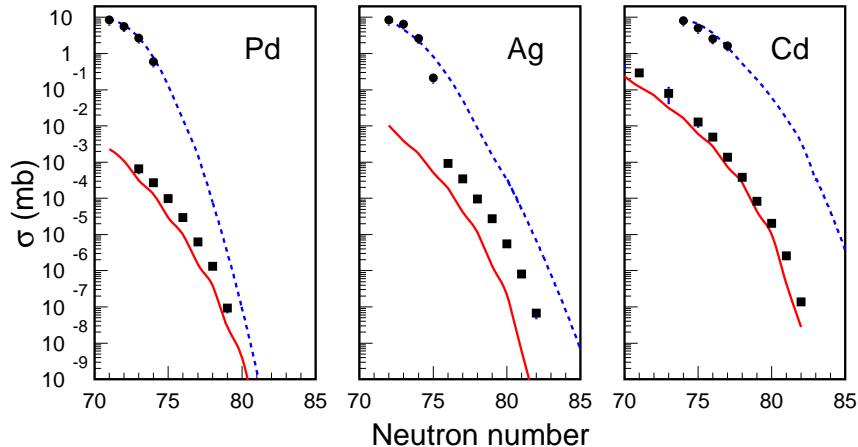,height=7cm}
\caption{Production cross sections of palladium, silver and cadmium isotopes produced in the fragmentation of $^{136}$Xe+Be at 1000 A MeV measured in this work (squares) and calculated with the code COFRA (solid line) compared to the productions obtained in the in-flight fission of $^{238}$U projectiles at 1000 A MeV impinging a lead target measured in \cite{Enq99} (circles) and calculated with the code SIMFIS \cite{Ben98} (dashed line).}
\label{fig_6}
  \end{center}
\end{figure*}

The alternatives to the fragmentation of $^{136}$Xe for the production of neutron-rich N=82 isotones below $^{132}$Sn are the in-flight fission of heavy projectiles, mainly $^{238}$U \cite{Ber97}, or a two-stage reaction scheme where highly volatile fission residues delivered by an ISOL facility, such as $^{132}$Sn, are re-accelerated and fragmented in a secondary target to produce neutron-rich nuclei by cold-fragmentation \cite{Hel03}. 

In-flight fission has extensively been investigated at the FRagment Separator facility at GSI where the isotopic production cross sections of fission residues have been measured in reactions induced by relativistic $^{238}$U \cite{Ber97,Eng97,Enq99,Ber03,Per06}, $^{208}$Pb \cite{Enq01,Enq02,Fer05} and $^{197}$Au \cite{Ben01} projectiles impinging on different target materials. This extensive data set allowed to investigate in detail the fission process in the large excitation-energy range populated in peripheral collisions between heavy ions at relativistic energies \cite{Jur,Ben02b,Ben06}, and formulate reaction models describing the isotopic composition of the final fission residues \cite{Ben98}.

Calculations for the expected production of N=82 isotones below $^{132}$Sn using a two-stage reaction scheme were presented in \cite{Hel03} while recently, an experiment was performed at GSI to measure such cross sections \cite{Per07}. In principle this two-stage reaction scheme is proposed to overcome the limited extraction efficiency of refractory elements, in which we are interested, in ISOL facilities. Since the results from this experiment are not yet available we limit the discussion in this section to the comparison of the production cross sections in fragmentation and fission reactions.

This comparison is shown in Fig. \ref{fig_6}, where we represent the production cross sections of palladium, silver and cadmium neutron-rich isotopes produced in the fragmentation of $^{136}$Xe projectiles at 1 A GeV impinging on a beryllium target measured in this work (squares) and calculated with the COFRA code (solid line). In the same figure we also depict the production of those nuclei in fission reactions induced by $^{238}$U projectiles at 950 A MeV impinging on a lead target measured in \cite{Enq99} (circles) and calculated with the Monte-Carlo-type SIMFIS code \cite{Ben98} (dashed line). In order to reach cross sections as low as picobarns the SIMFIS calculations were extrapolated using a polynomial function in exponential scale. For studying the production by fission, a lead target was chosen in order to enhance the production of very fissile slightly neutron-deficient uranium isotopes with moderate excitation energies by Coulomb excitation at relativistic energies \cite{Arm96}. 

As can be seen in the figure, fission of relativistic $^{238}$U projectiles, induced in a lead target, results in larger production cross sections of neutron-rich nuclei below $^{132}$Sn than the fragmentation of $^{136}$Xe projectiles. The difference with fragmentation decreases with the atomic number since we approach the region of symmetric fission (palladium), poorly populated by the electromagnetic-induced fission of $^{238}$U \cite{Arm96}.   

The final production rates of medium-mass neutron-rich nuclei in the next-generation radioactive-ion-beam facilities will depend not only on the production cross sections, which are investigated in this work but also on primary-beam intensities and target thicknesses, and on the technique used for the production; i.e. ISOL or in-flight with their different extraction or transmission efficiencies. In the case of ISOL facilities, the main technical difference affecting spallation reactions induced on a thick $^{238}$U target, is the extraction efficiency from the target. In principle, uranium composites present good release properties \cite{Rou02}. However, as already mentioned, the refractory nature of the elements of interest results in extremely low extraction efficiencies \cite{targi,Luk06,Rou06}. In this case, the two-stage scenario, where non-refractory neutron-rich fission residues like $^{132}$Sn can be easily extracted from a $^{238}$U target, post-accelerated, and then fragmented will allow the production of refractory neutron-rich nuclei below $^{132}$Sn with much higher yields.

In the case of in-flight facilities, the different kinematic properties of fission and fragmentation residues directly affects the transmission efficiencies of the reaction residues through the corresponding magnetic spectrometers. In present spectrometers such as the FRS, the difference in transmission can be up to a factor of 20 in favour of fragmentation. However, the large acceptance of the new-generation spectrometers, as BigRips in RIKEN \cite{Kub03} or the SuperFRS \cite{Gei03} at the future FAIR facility, reduces this difference in transmission down to a factor of 5. Therefore, in this case the final production rates will be determined to a much larger extent by the production cross sections. The comparison shown in Fig. \ref{fig_6} clearly shows that fission will be very competitive for the production of neutron-rich nuclei with atomic numbers between 60 and 48, and 44 and 32 populated by the asymmetric fission of $^{238}$U by the heavy and light fission fragments, respectively. However, the region of symmetric fission residues with atomic numbers between 45 and 47 will be better covered by the fragmentation of $^{136}$Xe.  

\section{Conclusion}

In this work we have investigated the production of medium-mass neutron-rich nuclei. For this purpose, an experiment was conducted at GSI to measure the production cross sections of neutron-rich projectile residues produced in the fragmentation of $^{136}$Xe projectiles at 1 A GeV in a beryllium target. Using the high-resolution magnetic spectrometer FRS we were able to identify more than 100 neutron-rich residual nuclei, produced in this reaction with cross sections as low as 100 pb, and to measure their production cross sections. Moreover, $^{125}$Pd was identified and its production cross section in fragmentation reactions was measured for the first time.

These measurements were used to benchmark two reaction codes describing the production cross sections of residual nuclei in fragmentation reactions, EPAX and COFRA. This analysis shows that the EPAX formula yields accurate predictions for the production cross sections of neutron-deficient nuclei and neutron-rich ones relatively close to the initial projectile in mass number. However this formula clearly overestimates the production cross sections of neutron-rich nuclei with four or more protons lost in the reaction. The code COFRA, which can only predict the production of residual nuclei with a large neutron excess, provides an overall good description of the production cross sections of the neutron-rich nuclei produced in this work. Only for nuclei very far in mass number from the projectile a moderate under-prediction of the production cross sections can be observed.

The present data and model calculations obtained for fragmentation reactions were also confronted to data and model calculations providing the production cross sections of similar nuclei but using fission reactions. This comparison shows that fission reactions are very competitive for the production of medium-mass neutron-rich nuclei within the atomic number intervals covered by the asymmetric fission of $^{238}$U. In the case of radioactive-ion-beam facilities using the in-flight technique, fragmentation of $^{136}$Xe can be used for the production of neutron-rich nuclei populating the region of symmetric fission (Z$\approx$ 44 to 47). In general, ISOL facilities suffer from the refractory nature of many of the N=82 neutron-rich isotones. In this case, the proposed two stage scenario for the production of those nuclei in the fragmentation of non-refractory fission residues such as $^{132}$Sn could be used  to overcome this limitation.

\begin{acknowledgments}

The authors are indebted to K.H. Behr, A. Br\"unle and K. Burkhard for their
technical support to this experiment. This work was partially supported  by the European Community under the FP6 ``Integrated Infrastructure Initiative EURONS'' contract n. RII3-CT-2004-506065 and the ``Research Infrastructure Action -Structuring the European Research Area EURISOL-DS'' contract n. 515768 RIDS, by the Spanish Ministry of Education and Science under grants FPA2005-00732 and FPA2007-62652 and the programme ``Ingenio 2010, Consolider CPAN'', and by ``Xunta de Galicia'' under grant PGIDT00PXI20606PM. The EC is not liable for any use that may be made of the information contained herein.

\end{acknowledgments}

\begin{appendix}

\small
\section{Measured cross sections} \label{appendA}

Production cross sections of fragmentation residues produced in collisions induced by 1000 A MeV $^{136}$Xe with beryllium. The number in parentheses represents the uncertainty referred to the corresponding last digits of the measured value.

\begin{table*}


\begin{tabular}{ccc}

\begin{tabular}{lr}
\hline
\hline
Isotope & $~$ $\sigma$(mb)\\ \hline
$^{136}$Cs & 0.60(7)\\
$^{135}$Cs & 1.55(17)\\
$^{134}$Cs & 2.59(29)\\
$^{133}$Cs & 3.22(36) \\
$^{132}$Cs & 3.16(35) \\
$^{135}$Xe & 109(12) \\
$^{134}$Xe & 51.1(56) \\
$^{133}$Xe & 31.7(35) \\
$^{132}$Xe & 28.8(32) \\
$^{131}$Xe & 19.9(22) \\
$^{130}$Xe & 16.7(18) \\
$^{129}$Xe & 13.3(15) \\
$^{136}$I & 3.6E-2(4) \\
$^{135}$I & 16.7(18) \\
$^{134}$I & 16.3(18) \\
$^{133}$I & 18.8(21)\\
$^{132}$I & 18.1(20) \\
$^{131}$I & 18.9(21) \\
$^{130}$I & 17.1(19) \\
$^{129}$I & 18.6(21) \\
$^{128}$I & 15.9(18) \\
$^{127}$I & 15.5(17) \\
$^{135}$Te & 2.7E-3(3) \\
$^{134}$Te & 0.30(3) \\
$^{133}$Te & 1.18(13) \\
$^{132}$Te & 2.55(12) \\
$^{130}$Te & 5.24(58) \\
$^{129}$Te & 6.97(78)\\
$^{128}$Te & 7.54(83) \\
$^{127}$Te & 9.60(99) \\
$^{126}$Te & 7.54(83) \\
$^{125}$Te & 10.9(12) \\
$^{134}$Sb & 3.09E-5(37) \\
$^{133}$Sb & 6.69E-3(73) \\
$^{132}$Sb & 4.49E-2(49) \\
$^{131}$Sb & 0.19(2) \\
$^{130}$Sb & 0.42(5)\\

\hline
\hline
\end{tabular}
$~~$
&
$~~$
\begin{tabular}{lr}
\hline
\hline
Isotope & $~$ $\sigma$(mb) \\ \hline
$^{129}$Sb & 0.84(9) \\
$^{127}$Sb & 2.22(25) \\
$^{126}$Sb & 3.02(33) \\
$^{125}$Sb & 4.10(45) \\
$^{124}$Sb & 5.31(59) \\
$^{123}$Sb & 6.25(69) \\
$^{132}$Sn & 1.19E-4(13) \\
$^{131}$Sn & 1.22E-3(13) \\
$^{130}$Sn & 7.29E-3(81) \\
$^{129}$Sn & 2.75E-2(30) \\
$^{128}$Sn & 7.45E-2(92) \\
$^{127}$Sn & 0.18(2) \\
$^{125}$Sn & 0.64(8) \\
$^{124}$Sn & 1.04(13) \\
$^{123}$Sn & 1.56(18)\\
$^{122}$Sn & 2.28(26) \\
$^{121}$Sn & 3.17(35) \\
$^{120}$Sn & 4.23(47) \\
$^{131}$In & 6.75E-6(77) \\
$^{130}$In & 5.35E-5(63) \\
$^{129}$In & 3.65E-4(41) \\
$^{128}$In & 1.78E-3(19) \\
$^{127}$In & 6.49E-3(72) \\
$^{126}$In & 1.89E-2(57) \\
$^{125}$In & 5.80E-2(75) \\
$^{124}$In & 0.13(7) \\
$^{122}$In & 0.38(5) \\
$^{121}$In & 0.63(7) \\
$^{120}$In & 1.03(12) \\
$^{119}$In & 1.58(18) \\
$^{118}$In & 2.24(25) \\
$^{130}$Cd & 1.39E-7(30) \\
$^{129}$Cd & 2.57E-6(31) \\
$^{128}$Cd & 2.06E-5(23) \\
$^{127}$Cd & 8.30E-5(97) \\
$^{126}$Cd & 3.77E-4(24) \\
$^{125}$Cd & 1.36E-3(16) \\
\hline
\hline
\end{tabular}

$~~$
&
$~~$

\begin{tabular}{lr}
\hline
\hline
Isotope & $~$ $\sigma$(mb)\\ \hline

$^{124}$Cd & 4.88E-3(55) \\
$^{123}$Cd & 1.29E-2(39) \\
$^{121}$Cd & 8.03E-2(40) \\
$^{119}$Cd & 0.29(4) \\
$^{118}$Cd & 0.44(5) \\
$^{117}$Cd & 0.83(10) \\
$^{116}$Cd & 1.18(13) \\
$^{115}$Cd & 1.78(20) \\
$^{128}$Ag & 6.77E-8(217) \\
$^{127}$Ag & 8.19E-8(116) \\
$^{126}$Ag & 5.47E-6(63) \\
$^{125}$Ag & 2.73E-5(33) \\
$^{124}$Ag & 9.61E-5(11) \\
$^{123}$Ag & 3.46E-4(41) \\
$^{122}$Ag & 9.28E-4(232) \\
$^{125}$Pd & 9.38E-8(271) \\
$^{124}$Pd & 1.34E-6(72) \\
$^{123}$Pd & 6.25E-6(72) \\
$^{122}$Pd & 2.96E-5(20) \\
$^{121}$Pd & 9.92E-5(12) \\
$^{120}$Pd & 2.67E-4(53) \\
$^{119}$Pd & 6.61E-4(198) \\
$^{122}$Rh & 2.05E-7(47) \\
$^{121}$Rh & 1.49E-6(19) \\
$^{120}$Rh & 5.36E-6(62) \\
$^{119}$Rh & 1.86E-5(22) \\
$^{118}$Rh & 5.26E-5(63) \\
$^{117}$Rh & 1.36E-4(32) \\
$^{119}$Ru & 2.72E-7(52) \\
$^{118}$Ru & 1.42E-6(18) \\
$^{117}$Tc & 5.28E-8(208) \\
$^{116}$Tc & 2.23E-7(46) \\
$^{115}$Tc & 3.43E-6(45) \\
$^{113}$Mo & 2.59E-7(51) \\
$^{111}$Nb & 4.08E-8(172) \\
$^{110}$Nb & 2.51E-7(51) \\
$^{108}$Zr & 6.05E-8(224) \\

\hline
\hline
\end{tabular}

\end{tabular}


\end{table*}


\end{appendix}


\end{document}